\documentclass[aps,prx,preprint,superscriptaddress]{revtex4-2}
\usepackage{latexsym}
\usepackage{graphicx}
\begin{document}

\title{High-sensitivity of initial SrO growth on the residual resistivity in epitaxial thin films of SrRuO$_3$ on SrTiO$_3$(001)}


%\author{}

%\affiliation{}
\affiliation{Institute of Physics, Academia Sinica, Nankang, Taipei 11529, Taiwan}
\affiliation{National Synchrotron Radiation Research Center, Hsinchu 30076, Taiwan}
\affiliation{Nano Science and Technology, Taiwan International Graduate Program, Academia Sinica and National Taiwan university.}
\author{Uddipta Kar}\thanks{These authors contributed equally to the work.}
\affiliation{Institute of Physics, Academia Sinica, Nankang, Taipei 11529, Taiwan}
\affiliation{Nano Science and Technology, Taiwan International Graduate Program, Academia Sinica and National Taiwan university.}
\author{Akhilesh Kr. Singh}\thanks{These authors contributed equally to the work.}
\affiliation{Institute of Physics, Academia Sinica, Nankang, Taipei 11529, Taiwan}
\author{Song Yang}
\affiliation{National Synchrotron Radiation Research Center, Hsinchu 30076, Taiwan}
\author{Chun-Yen Lin}
\affiliation{National Synchrotron Radiation Research Center, Hsinchu 30076, Taiwan}
\author{Bipul Das}
\affiliation{Institute of Physics, Academia Sinica, Nankang, Taipei 11529, Taiwan}
\author{Chia-Hung Hsu}\email{chsu@nsrrc.org.tw}
\affiliation{National Synchrotron Radiation Research Center, Hsinchu 30076, Taiwan}
\author{Wei-Li Lee}\email{wlee@phys.sinica.edu.tw}
\affiliation{Institute of Physics, Academia Sinica, Nankang, Taipei 11529, Taiwan}

\date{\today}

\begin{abstract}
The growth of SrRuO$_3$ (SRO) thin film with high-crystallinity and low residual resistivity (RR) is essential to explore its intrinsic properties. Here, utilizing the adsorption-controlled growth technique, the growth condition of initial SrO layer on TiO$_2$-terminated SrTiO$_3$ (STO) (001) substrate was found to be crucial for achieving a low RR in the resulting SRO film grown afterward. The optimized initial SrO layer shows a \textit{c}(2 $\times$ 2) superstructure that was characterized by electron diffraction, and a series of SRO films with different thicknesses ($t$s) were then grown. The resulting SRO films exhibit excellent crystallinity with orthorhombic-phase down to $t \approx$ 4.3 nm, which was confirmed by high resolution X-ray measurements. From X-ray azimuthal scan across SRO orthorhombic (02$\pm$1) reflection, we uncover four structural domains with a dominant domain of orthorhombic SRO [001] along cubic STO [010] direction. The dominant domain population depends on $t$, STO miscut angle ($\alpha$), and miscut direction ($\beta$), giving a volume fraction of about 92 $\%$ for $t \approx$ 26.6 nm and $(\alpha, \beta) \approx$ (0.14$^{\rm o}$, 5$^{\rm o}$). On the other hand, metallic and ferromagnetic properties were well preserved down to \textit{t} $\approx$ 1.2 nm. Residual resistivity ratio (RRR = $\rho({\rm 300 K})$/$\rho({\rm 5K})$) reduces from 77.1 for \textit{t} $\approx$ 28.5 nm to 2.5 for \textit{t} $\approx$ 1.2 nm, while $\rho({\rm 5K})$ increases from 2.5 $\mu\Omega$cm for \textit{t} $\approx$ 28.5 nm to 131.0  $\mu\Omega$cm for \textit{t} $\approx$ 1.2 nm. The ferromagnetic onset temperature ($T'_{\rm c}$) of around 151 K remains nearly unchanged down to \textit{t} $\approx$ 9.0 nm and decreases to 90 K for \textit{t} $\approx$ 1.2 nm. Our finding thus provides a practical guideline to achieve high crystallinity and low RR in ultra-thin SRO films by simply adjusting the growth of initial SrO layer. 

\end{abstract} 

\maketitle
\section{Introduction}
The orthorhombic SRO hosts a number of intriguing physical properties, such as ferromagnetism with $T_{\rm c} \approx$ 160 K \cite{kos2012}, Fermi-liquid behavior \cite{cap2002}, magnetic monopole \cite{fan2003}, and Weyl fermions \cite{ito2016,jen2019}. The growth of SRO in thin-film form may open up possibilities to further tune its unusual physical properties by strain and finite-size effects. Extensive efforts have been carried out previously to grow high-crystalline SRO films on various substrates \cite{kos2012,nai2018,mac1998,roy2015,shi2011}, where different transport and magnetic properties were found as compared to its bulk form. On the other hand, the issue of the critical thickness for the structural and magnetic phase transitions in ultra-thin SRO films remains a debatable issue, where the strain and substrate symmetry play important roles \cite{he2010,zay2008,her2015}. Theoretical outcomes infer the ferromagnetic and metallic phases in the SRO films on STO down to a thickness of about 1 nm \cite{maha2009,Si2015}, while experimental resolutions showed more scattered and inconclusive results due to the difficulty on maintaining high crystallinity in ultra-thin SRO films \cite{toyota2005,xia2009,cha2011,bos2019}.

In the past, SRO films grown using sintered oxide targets by sputtering or pulsed laser deposition showed relatively low RRRs ($<$ 10) \cite{cha2009,dab2004,gan1998,lee2017,sie2007}. On the other hand, thick SRO films grown by electron beam evaporation technique turned out to give much higher RRR ($>$ 60) \cite{roy2015,mac1998} that is approaching the value in a bulk single crystal. Such a large difference in RRR suggests a high sensitivity of the SRO stoichiometry on growth parameters, where the volatility of ruthenium oxide and thus cation deficiency turn out to be a major problem \cite{sch2016,kos2012}. In this respect, an adsorption-controlled growth technique was developed for thin-film growth of various oxides, such as PbTiO$_3$, EuO, BaSnO$_3$, and LuFe$_2$O$_4$ \cite{ulb2008,pra2017,bro2012,the1998}. We noted that similar technique was first introduced in the growth of GaAs films \cite{art1968}. More recently, the growth of high-quality chalcogenide thin films also relied on this approach \cite{chen2011}. In the adsorption-controlled technique, the flux ratio of source materials and the growth temperature are key parameters to grow a stoichiometric film, and a thermodynamic phase diagram can be constructed to reveal the proper growth window for a particular structural phase. 

For the adsorption-controlled growth of orthorhombic SRO films on STO, a growth window of ozone partial pressure of around 3 $\times$ 10$^{-6}$ Torr and a growth temperature ranging from about 500 $^{\rm o}$C to 800 $^{\rm o}$C was reported previously \cite{nai2018}. Above 800 $^{\rm o}$C, other phases of Sr$_4$Ru$_3$O$_{10}$ and Sr$_2$RuO$_4$ become more thermodynamically favorable. Within the growth window, the supplied Ru flux forms a volatile RuO$_{\rm x}$ and desorbs from the film surface. The SRO growth will happen when the RuO$_{\rm x}$ combine with the SrO, and the growth rate is thus controlled by the Sr flux. With the appropriate flux ratio of Ru/Sr, the film's stoichiometry can be thermodynamically self-regulated, resulting in a high-quality and single phase SRO films on STO. However, for the adsorption-controlled growth technique, the questions in regard to the initial growth condition \cite{hon2005,rij2004,san2003} and its influence on the follow-up SRO growth are still not well understood. In this work, we used an oxide-MBE and adopted the adsorption-controlled growth technique to grow SRO films with different $t$s on TiO$_2$-terminated STO (001)$_{\rm c}$ substrates, where the subscript c refers to a cubic phase. From reflection high-energy electron diffraction (RHEED) and low energy electron diffraction (LEED) analyses, we found that an optimized initial SrO layer gave a \textit{c}(2 $\times$ 2) superstructure, which turned out to be a prerequisite for excellent crystallinity and low residual resistivity in the resulting SRO films. The SRO films grown with optimized initial SrO layer showed an orthorhombic-phase down to \textit{t} $\approx$ 4.3 nm. In addition, the structural domains in our SRO films were investigated by performing X-ray azimuthal scans across the SRO (02$\pm$1)$_{\rm o}$ reflections, where the subscript o refers to an orthorhombic phase. We remark that the films grown with the optimized initial SrO layer give a significant reduction in RR as compared to films grown with unoptimized initial SrO layer.

\section{Results}
\begin{figure}
\includegraphics[width=\linewidth]{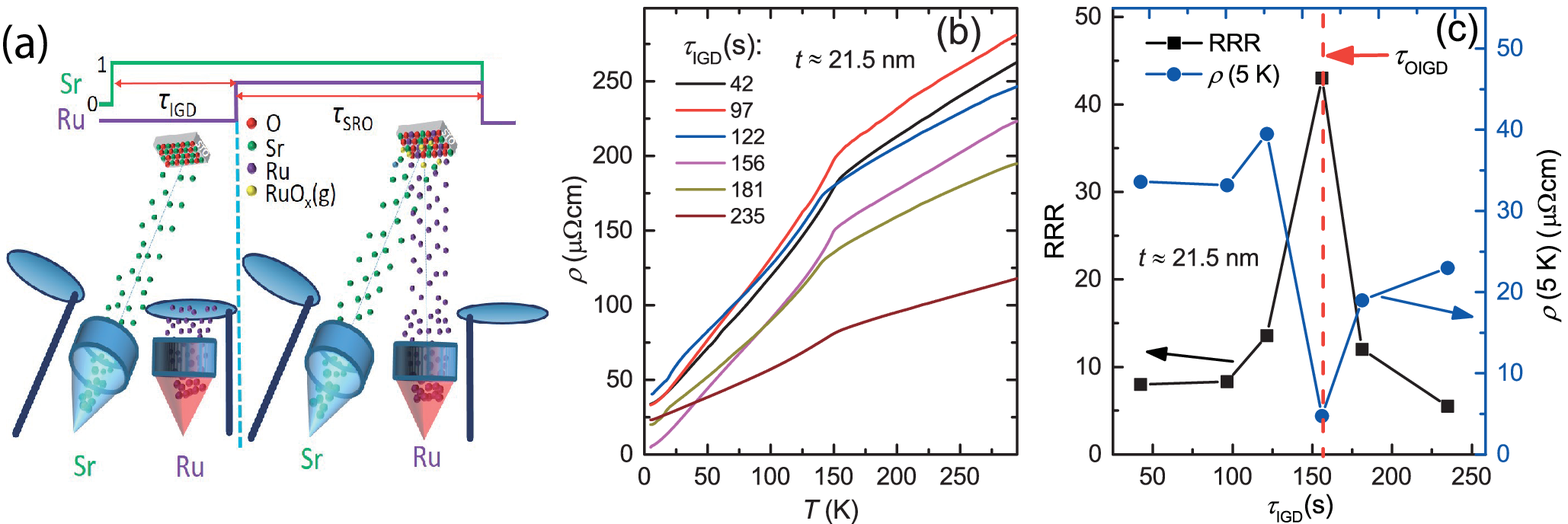}
  \caption {The influence of initial SrO growth condition on $\rho$(5 K) and RRRs of the SRO films on STO. (a) A schematic cartoon illustrating the growth of a SRO, where 0 and 1 refer to the shutter close and open, respectively. Ru shutter was opened when Sr flux had been turned on for a period of $\tau_{\rm IGD}$. (b) For $t$ $\approx$ 21.5 nm, the temperature dependent resistivity of SRO films grown with different $\tau_{\rm IGD}$ values. (c) shows a nonmonotonic variation of the corresponding RRR and $\rho$(5 K) with respect to $\tau_{\rm IGD}$, revealing an optimum vlaue of $\tau_{\rm IGD}$ = $\tau_{\rm OIGD}$ for achieving lowest $\rho$(5 K) and thus highest RRR in SRO films. }
  \label{Fig1}
\end{figure}

 Figure \ref{Fig1}(a) shows a schematic of the SRO film growth, where the operations of the Ru and Sr cell shutters are illustrated. Sr shutter was first opened for a certain initial growth duration ($\tau_{\rm IGD}$) for the growth of the initial SrO layer on a STO, and then Ru shutter was opened for the subsequent growth of SRO film. The resulting thickness ($t$) of SRO film can be well controlled by the Ru shutter opening time of $\tau_{\rm SRO}$. Figure \ref{Fig1}(b) shows the temperature dependent resistivity ($\rho$) with different $\tau_{\rm IGD}$ values for \textit{t} $\approx$ 21.5 nm films. A practical $T^2$ dependence in $\rho(T)$ was found for all films in low temperature regime, indicating a Fermi liquid behavior as expected \cite{cap2002}. For convenience, we use $\rho$ at $T$ = 5 K ($\rho$(5 K)) as a measure for RR of the SRO films in the following discussions. The extracted RRRs and $\rho$(5 K) from the $\rho(T)$ curves show nonmonotonic variations with $\tau_{\rm IGD}$ as plotted in Fig. \ref{Fig1}(c), where a maximum RRR of about 43.0 and a lowest $\rho$(5 K) of about 4.7 $\mu\Omega$cm were achieved for the film grown with an optimum $\tau_{\rm OIGD} \approx $156 s. Remarkably, $\rho$(5 K)(RRR) becomes higher(lower) by nearly an order of magnitude for films grown with the condition of $\tau_{\rm IGD} \neq \tau_{\rm OIGD}$. 

\begin{figure}
  \includegraphics[width=\linewidth]{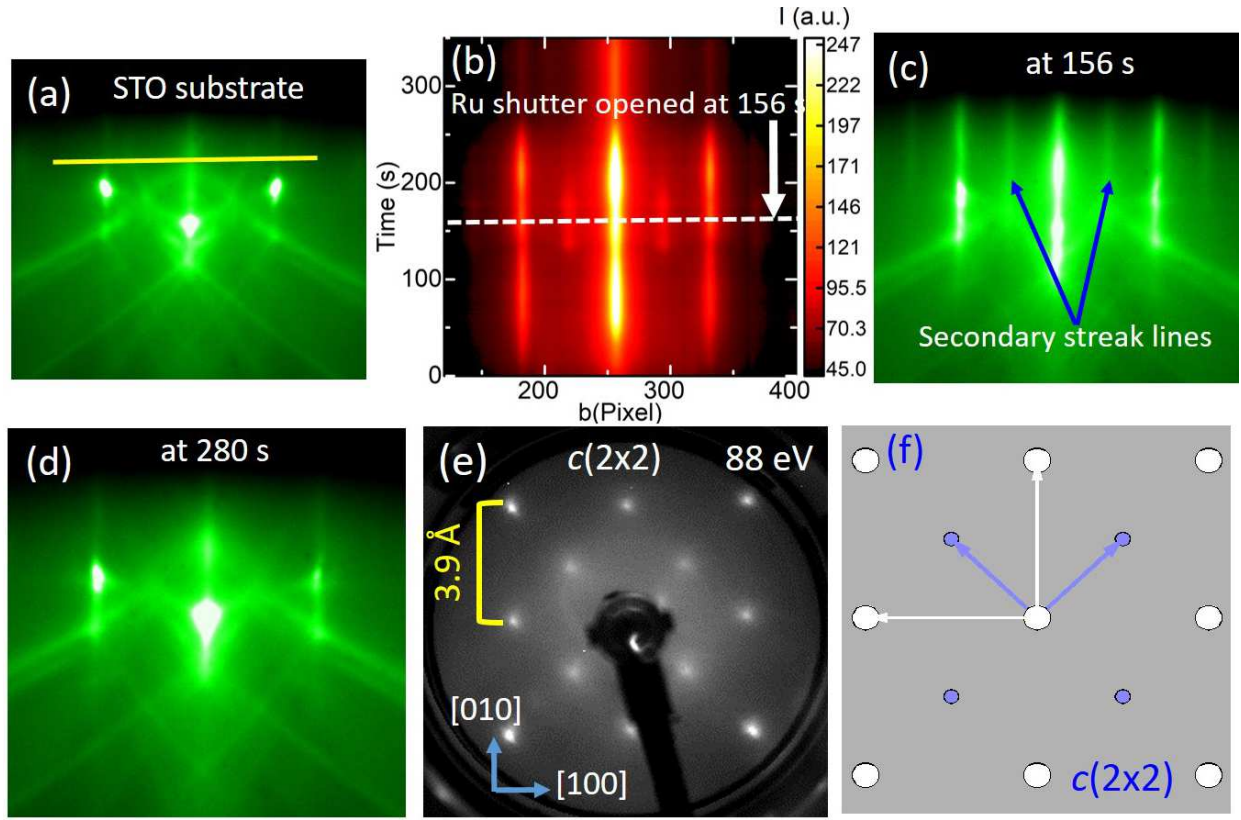}
  \caption{Structural evolution during the growth of SRO. (a) RHEED pattern of the TiO$_2$-terminated STO (001)$_{\rm c}$ substrate with electron beam along STO [110]$_{\rm c}$ direction. (b) Time evolution of RHEED intensity along the yellow line shown in (a). After supplying Sr, the spot-like RHEED pattern from the STO substrate started to evolve into a streak-line pattern, and secondary streak-lines then appeared between the main streak lines. The white dashed line in (b) represents the optimum time for opening of Ru shutter, and the corresponding RHEED pattern is shown in (c), where the pronounced secondary streak-lines are observed. (d) The RHEED pattern after the growth of about 0.4 nm of SRO (at the time of 280 s), and its pattern and intensity remain nearly unchanged with subsequent SRO growth, indicating a step-flow type growth process. (e) shows the LEED pattern of an optimum initial SrO layer on STO substrate with a beam energy of 88 eV, which agrees well with the simulated LEED pattern shown in (f) with a $c$(2 $\times$ 2) superstructure. }
  \label{Fig2}
\end{figure}

In order to know the structural evolution, we carefully monitored the initial growth using an \textit{in-situ} RHEED. Figure \ref{Fig2}(a) displays the RHEED pattern of the STO substrate along [110]$_{\rm c}$ direction at 700 $^{\rm o}$C \cite{tsa2017}. The time evolution of the RHEED intensity profile across the solid line in Fig. \ref{Fig2}(a) is shown in Fig. \ref{Fig2}(b). Upon opening Sr cell shutter, the RHEED image transformed from a spot-like pattern into a streak-line pattern. The secondary streak-lines started appearing between the primary streak-lines after $\tau_{\rm IGD} \approx$ 118 s. Figure \ref{Fig2}(c) shows the RHEED pattern right before opening the Ru shutter, where the pronounced secondary streak-lines were visible between the main streak-lines. After opening the Ru shutter at $\tau_{\rm OIGD} \approx$ 156 s as marked by the white dashed line in Fig. \ref{Fig2}(b), the secondary streak-lines gradually disappeared and the intensity of the primary streak-lines also reduced. After about 0.4 nm growth of SRO, the primary streak-lines transformed back to the spot-like feature as shown in Fig. \ref{Fig2}(d), indicating the structural transformation of the initial SrO layer due to the adsorption and incorporation of RuO$_{\rm x}$, and the RHEED pattern remained nearly unchanged after then. SRO films grown with the $\tau_{\rm IGD} < \tau_{\rm OIGD}$ followed the island-type growth. In contrast, for $\tau_{\rm IGD} > \tau_{\rm OIGD}$, the streak-line feature from the initial SrO layer remains nearly unchanged after opening the Ru shutter for subsequent growth of the SRO film.

To further explore the surface structure prior the SRO growth, we grew a SrO layer on STO with $\tau_{\rm IGD} = \tau_{\rm OIGD}$ at growth temperature of 700 $^{\rm o}$C. The secondary streak-lines were found to be stable while cooling down to room temperature, and then the sample was transferred under an ultra-high vacuum to a LEED chamber for surface structural characterizations. Figure \ref{Fig2}(e) shows the LEED pattern with a beam energy of 88 eV. The primary spots come from the cubic STO substrate with a lattice spacing of 3.91 $\rm\AA$. The secondary spots appeared along the lateral $<$110$>$ directions halfway between the main spots. The simulated LEED pattern \cite{her2014} with a \textit{c}(2 $\times$ 2) structure as shown in Fig. \ref{Fig2}(f) exhibits a close agreement with the pattern we observed. These observations confirmed a \textit{c}(2 $\times$ 2) superstructure on the surface of the initial SrO layer \cite{jal2009}.

\begin{figure}
 \includegraphics[width=\linewidth]{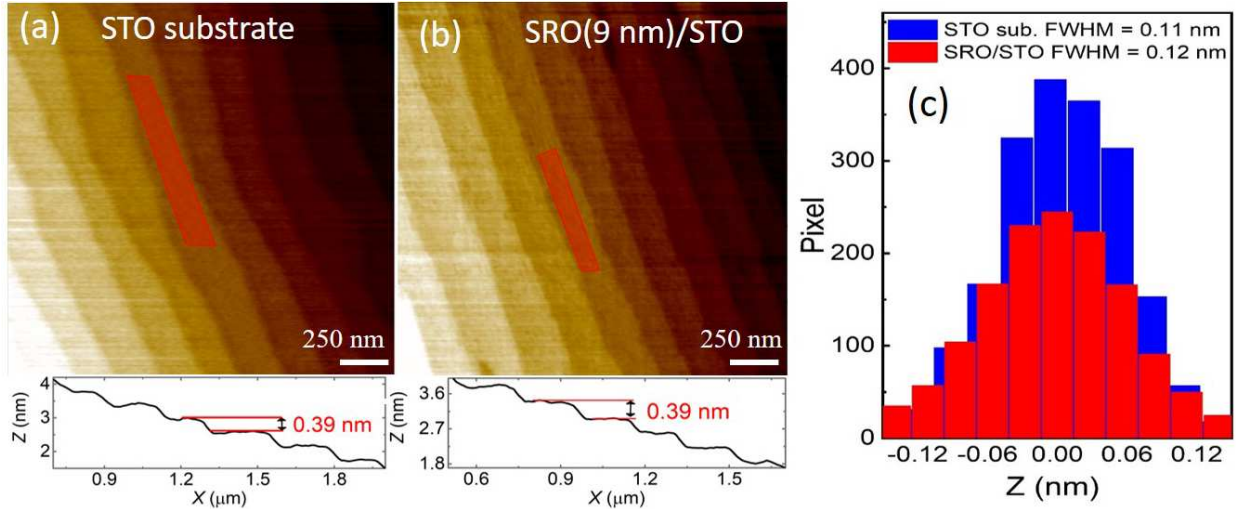}
  \caption{Surface morphology of a SRO film on STO before and after the growth. (a) AFM image of a TiO$_2$-terminated STO substrate. (b) AFM image of a SRO film with \textit{t} $\approx$ 9 nm. Lower panel of (a) and (b) show the line profile of the terraces, where the step-height of around 0.39 nm corresponds to one unit-cell height of STO. (c) Histogram of height distribution within the shaded region of (a) and (b). The film roughness remains nearly the same as that of the bare substrate.}
\label{Fig3}
\end{figure}

By using the same condition of $\tau_{\rm IGD} = \tau_{\rm OIGD}$, we grew a series of SRO(\textit{t}) films on STO with different \textit{t} ranging from 1.2 nm to 28.5 nm. Figures \ref{Fig3} (a) and (b) show the atomic force microscope (AFM) images before and after the growth of a \textit{t} $\approx$ 9 nm  film, respectively. The height profiles shown in the lower panel of Figs. \ref{Fig3}(a--b) indicate the well preservation of atomic steps (height $\approx$ 0.4 nm) after the SRO film growth. The terrace width of the bare STO substrate was around 150 to 600 nm, which gives a miscut angle of around $\alpha \approx$ 0.15$^{\rm o}$ to 0.04$^{\rm o}$. Figure \ref{Fig3}(c) shows the height-histogram of the shaded area in the Figs. \ref{Fig3}(a--b). The average surface roughness of around 0.12 nm after the growth remains nearly the same as that of the STO substrate. However, we did notice some well-separated random clusters on film surface. Using an energy dispersive X-ray analyzer (SEM-EDX), we confirmed that those clusters were composed of RuO$_{\rm x}$ \cite{kos2012}. Those random RuO$_{\rm x}$ clusters on the film surface are likely coming from the excess supplied Ru during the growth. The density of the RuO$_{\rm x}$ clusters can be minimized by reducing the fluxes of both Sr and Ru, while keeping the same flux ratio.

\begin{figure}
\includegraphics[width=14 cm]{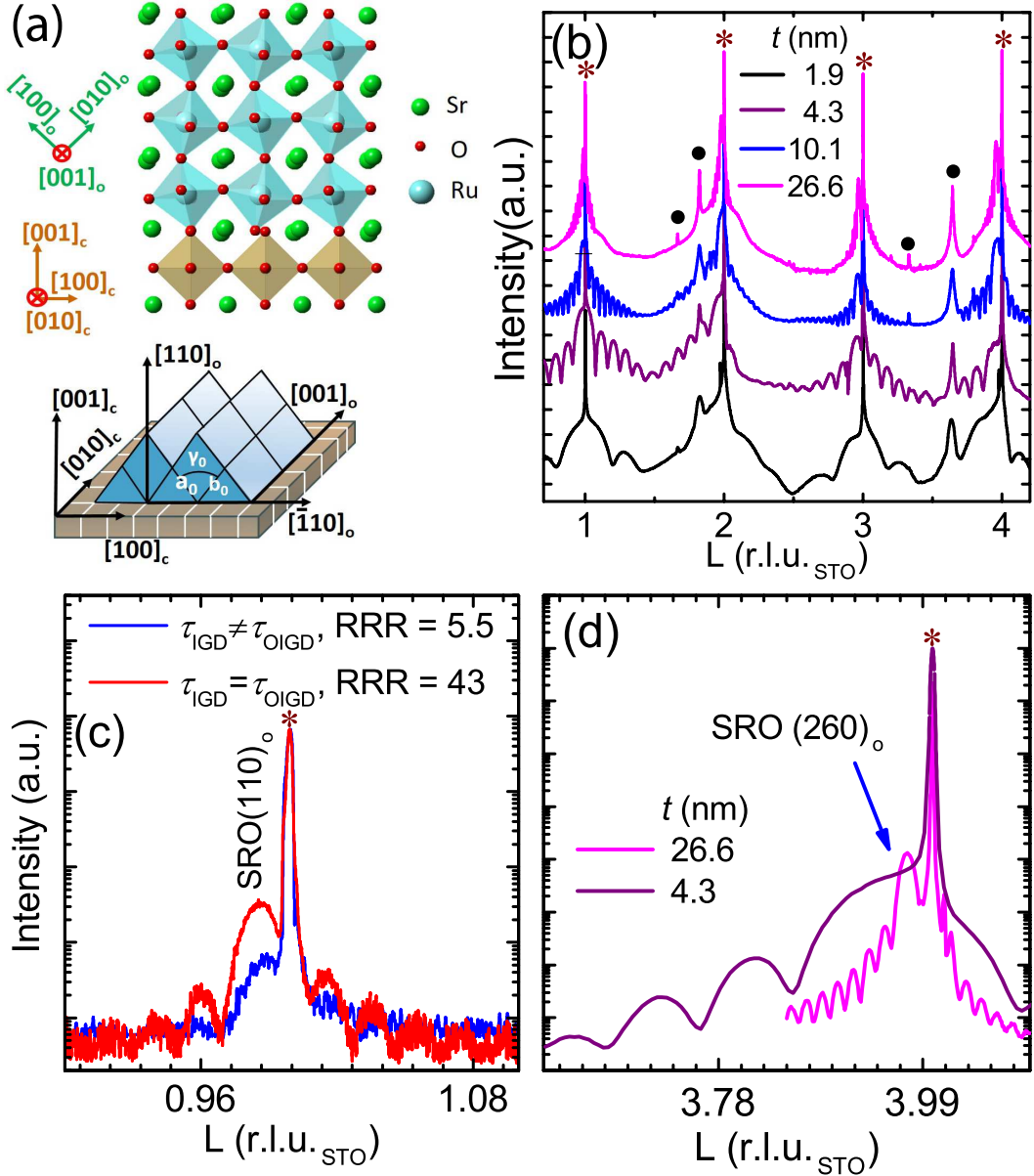}
  \caption{Structural investigation of SRO films on STO using XRD. (a) displays the tilting of the RuO$_6$ octahedra of an orthorhombic SRO film on a cubic STO. (b) CTR of the SRO(\textit{t})/STO films. The presence of fringes around the Bragg peaks revealed the excellent crystalline quality of the SRO films and sharp interfaces. The peaks marked by solid spheres are attributed to the random RuO$_{\rm x}$ clusters present on the film surface. (c) Radial scan along surface normal of \textit{t} $\approx$ 21.5 nm films grown under $\tau_{\rm IGD} = \tau_{\rm OIGD}$ and $\tau_{\rm IGD} \neq \tau_{\rm OIGD}$. The pronounced Laue oscillations appeared only for the film grown with $\tau_{\rm IGD} = \tau_{\rm OIGD}$. (d) L-scan of the off-normal SRO (260)$_{\rm o}$ reflection for SRO films with \textit{t} $\approx$ 26.6 and 4.3 nm. Pronounced fringes with a period nearly the same as that of corresponding specular rod reveal the excellent lateral crystalline quality of the SRO films.}
  \label{Fig4}
\end{figure}

Figure \ref{Fig4}(a) shows a schematic for the crystalline orientation of orthorhombic SRO on a cubic STO (001)$_{\rm c}$ substrate. Figure \ref{Fig4}(b) displays STO (00n)$_{\rm c}$, where n is an integer, crystal truncation rods (CTRs) of the samples with various thicknesses. The abscissa L represents the momentum transfer along surface normal and is in unit of STO reciprocal lattice unit (r.l.u.) with a value of 1.609 $\rm\AA^{-1}$. SRO (nn0)$_{\rm o}$ reflections were found centered at slightly low-L side of STO (00n)$_{\rm c}$ reflections as expected for the SRO (110)$_{\rm o}$ oriented films \cite{cha2011,nai2018}. Moreover, the presence of intensity oscillations around the SRO Bragg reflection manifests the excellent crystallinity and sharp interfaces of the SRO films. The oscillation periods of 0.0149, 0.0373, 0.0883, and 0.216 r.l.u.$_{\rm STO}$ give the crystalline layer thicknesses of 26.2, 10.5, 4.3, and 1.9 nm, respectively. We noted two unidentified peaks (black solid circles in Fig. \ref{Fig4}(b)) at L = 1.82 and 3.64 r.l.u.$_{\rm STO}$ for all samples. Two additional minor peaks were also observed at L = 1.66 and 3.33 r.l.u.$_{\rm STO}$. These additional peaks belong to the RuO$_{\rm x}$ clusters present on the film surface \cite{kos2012,jos2017}. Figure \ref{Fig4}(c) shows radial scans along surface normal for SRO films with \textit{t} $\approx$ 21.5 nm grown with $\tau_{\rm IGD} = \tau_{\rm OIGD}$ and $\tau_{\rm IGD} \neq \tau_{\rm OIGD}$. Pronounced thickness fringes appeared near the SRO (110)$_{\rm o}$ Bragg peak for the film with $\tau_{\rm IGD} = \tau_{\rm OIGD}$. But for the $\tau_{\rm IGD} \neq \tau_{\rm OIGD}$ case, no fringes were observed, and the Bragg peak intensity was also weaker. This notable difference revealed a much better crystallinity of the SRO film grown with  $\tau_{\rm IGD} = \tau_{\rm OIGD}$. We further performed CTR measurements of various off-normal SRO reflections to examine the orientation and crystalline quality of films along the lateral directions. Figure \ref{Fig4}(d) shows L-scans across  the STO (204)$_{\rm c}$ reflection for SRO films with \textit{t} $\approx$ 26.6 nm and 4.3 nm. Both samples exhibit pronounced SRO Bragg peaks centered at $\approx$ 3.97 r.l.u.$_{\rm STO}$ and intensity oscillations, and the oscillation's period agrees well with that measured from the STO (00n)$_{\rm c}$ CTRs, further manifesting the excellent 3D crystalline quality of the SRO films along both normal and lateral directions. 

\begin{figure}
  \includegraphics[width=14 cm]{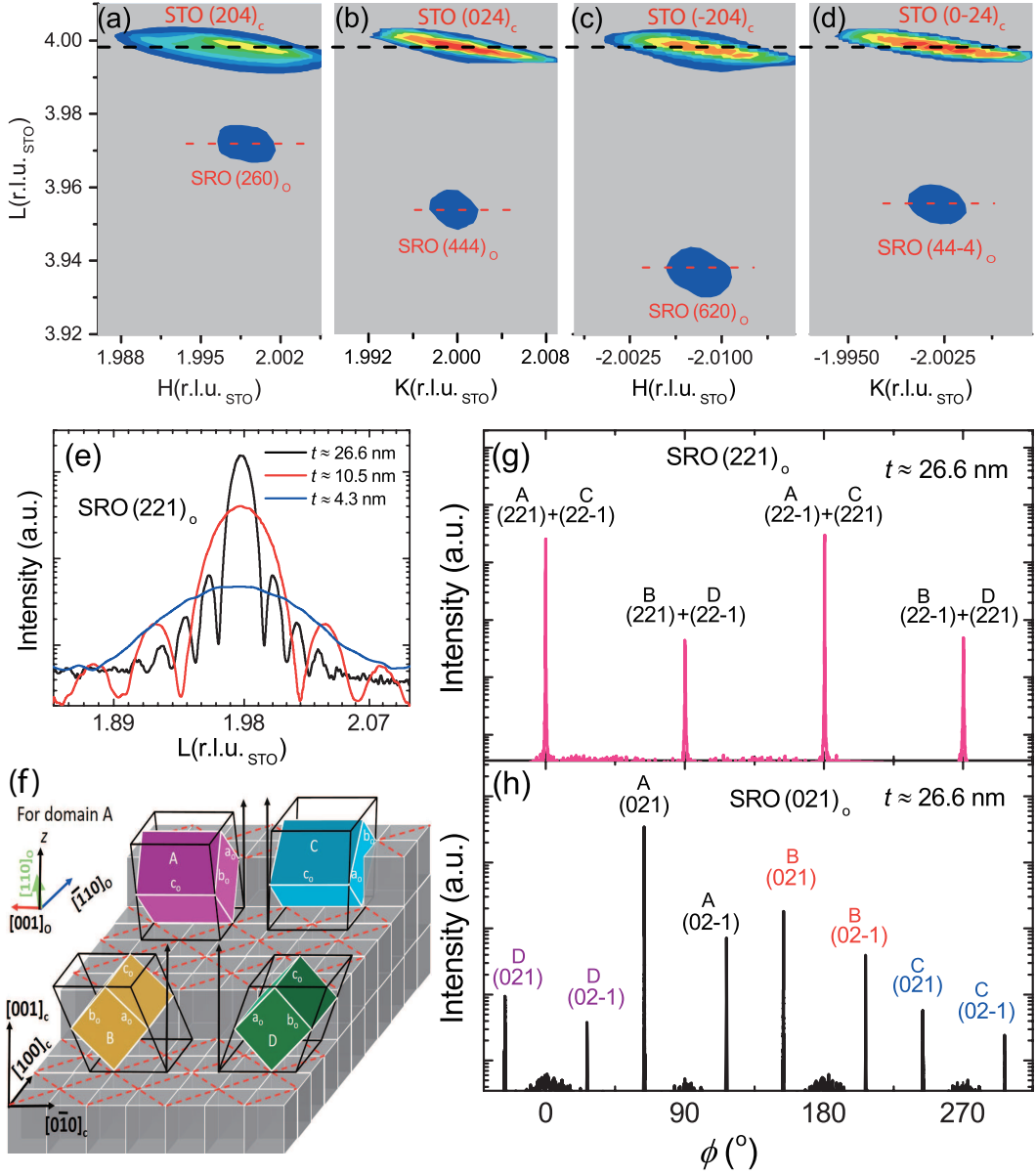}
  \caption{ Structural-phase determination of the SRO(\textit{t})/STO films. (a) RSM for a SRO film with \textit{t} $\approx$ 26.6 nm across (a) STO (204)$_{\rm c}$, (b) STO (024)$_{\rm c}$, (c) STO (-204)$_{\rm c}$, and (d) STO (0-24)$_{\rm c}$ reflections. (e) The L-scans of the SRO (221)$_{\rm o}$ refection for different \textit{t}s. The appearance of SRO (221)$_{\rm o}$ peaks infers the orthorhombic-phase down to $t$ $\approx$ 4.3 nm. (f) An illustration of the growth orientations for the four domains A, B, C, and D. Red dashed lines show the $c$(2 $\times$ 2) superstructure for initial SrO layer. (g) and (h) are the $\phi$-scans across the SRO (221)$_{\rm o}$ and SRO (021)$_{\rm o}$ reflections, respectively, indicating the coexistence of both 90$^{\rm o}$ and 180$^{\rm o}$ oriented domains in the SRO film.}
  \label{Fig5}
\end{figure}

We further moved on to the structural-phase evolution with respect to $t$ for films grown with $\tau_{\rm IGD}$ = $\tau_{\rm OIGD}$. Figures \ref{Fig5}(a--d) display the reciprocal space maps (RSMs) near the STO (204)$_{\rm c}$, (024)$_{\rm c}$, (-204)$_{\rm c}$ and (0-24)$_{\rm c}$ reflections for the sample of $t$ $\approx$ 26.6 nm. The intense peaks at L = 4 r.l.u.$_{\rm STO}$ correspond respectively to the STO (204)$_{\rm c}$, (024)$_{\rm c}$, (-204)$_{\rm c}$, and (0-24)$_{\rm c}$ reflections. The peaks at L = 3.973, 3.953, 3.938, and 3.955 r.l.u.$_{\rm STO}$ belong to the SRO (260)$_{\rm o}$, (444)$_{\rm o}$, (620)$_{\rm o}$, and (44-4)$_{\rm o}$ reflections, respectively. The apparent difference in the L values of the SRO (260)$_{\rm o}$ and SRO (620)$_{\rm o}$ reflections provided a distinct evidence for the orthorhombic phase \cite{cha2011,vai2007}. However, because of the similar unit cell size between tetragonal and orthorhombic phases of SRO, where the difference in lattice constants is less than 1$\%$, diffraction peaks of the two phases are always nearby. Consequently, peak indexing and phase identification from similar RSMs become practically impossible for films thinner than $\approx$ 10 nm, because the SRO peaks are so broad along L due to finite size effect and the peaks associated with different rotational domains overlap seriously. Hence, we chose the orthorhombic-specific reflections, such as SRO (221)$_{\rm o}$ and (021)$_{\rm o}$, as the signatures to identify orthorhombic phase. Originated from the tilt of the RuO$_6$ octahedra, those reflections are allowed in the orthorhombic phase but forbidden in the tetragonal phase \cite{cha2011}. Figure \ref{Fig5}(e) shows the thickness-dependent L-scans across the SRO (221)$_{\rm o}$ reflection, which was adopted as the signature of the orthorhombic phase. The peak width increases monotonically with decreasing $t$ from 26.6 to 4.3 nm, accompanied by increasing fringe period. On the other hand, no SRO (221)$_{\rm o}$ or (021)$_{\rm o}$ reflections were found for samples of $t$ $\approx$ 3.0 nm or thinner. These results clearly demonstrate that the SRO films transform from orthorhombic to tetragonal phase as the layer thickness reduces below $\approx$ 4.0 nm. The lattice parameters of the orthorhombic SRO derived by fitting the angular positions of at least four reflections for each sample are $a_{\rm o}$ = 5.584 $\pm$ 0.008 \AA, $b_{\rm o}$ = 5.540  $\pm$ 0.005 \AA, $c_{\rm o}$ = 7.810 $\pm$ 0.016 \AA, and  $\gamma_{\rm o}$ = 89.43$^{\rm o}$ $\pm$ 0.16$^{\rm o}$. The slight deviation of $\gamma_{\rm o}$ from 90$^{\rm o}$ is a result of the strain due to lattice mismatch with the STO substrate. Because the uncertainty given by weak peak intensity and broad peak width of the reflections of ultra-thin SRO films, no obvious trend in the variation of lattice parameters was concluded. 
 
Figure \ref{Fig5}(g) shows the azimuthal $\phi$-scan of the SRO (221)$_{\rm o}$ reflection for the $t$ $\approx$ 26.6 nm film grown on an on-axis substrate with ($\alpha, \beta$) $\approx$ (0.14$^{\rm o}$, 5$^{\rm o}$). Four evenly spaced pronounced peaks with alternating peak intensities were observed, suggesting the presence of at least two 90$^{\rm o}$ rotational domains. Nevertheless, there are four possible rotational variants for (110) oriented orthorhombic SRO domains as illustrated schematically in Fig. \ref{Fig5}(f). Because of the tiny difference between lattice constants \textit{a}$_{\rm o}$ and \textit{b}$_{\rm o}$ of orthorhombic SRO, the SRO (221)$_{\rm o}$ reflection of one domain overlaps with the SRO (22-1)$_{\rm o}$ reflection of the domain which is rotated 180$^{\rm o}$ against surface normal from the former one. We cannot differentiate the population between the two domains which are rotated 180$^{\rm o}$ with respect to each other from the SRO (221)$_{\rm o}$ $\phi$-scan. According to the selection rules of orthorhombic SRO with \textit{Pbnm} space group, the SRO (20$\pm$1)$_{\rm o}$ reflections are forbidden but the (02$\pm$1)$_{\rm o}$ pair are allowed, which are about 53.1$^{\rm o}$ apart azimuthally. The (02$\pm$1)$_{\rm o}$ pair associated with the four (110)-oriented rotational domains are offset by a multiple of 90$^{\rm o}$ in azimuthal angle, well separated from each other, and thus can be employed to determine the population of the four rotational domains. Figure \ref{Fig5}(h) illustrates the azimuthal scan across SRO (02$\pm$1)$_{\rm o}$ reflections of a SRO film with $t \approx$ 26.6 nm. With the direction of $\phi$ = 0 assigned to align with the STO [100]$_{\rm c}$ direction, the peak locations agree well with calculated $\phi$ angles for the four domains shown in Fig. 5(f). The intensity difference between the (021)$_{\rm o}$ and (02-1)$_{\rm o}$ peaks associated with the same domain may be attributed to X-ray foot print on the irregular sample shape, and the weak broad peaks in the middle of each (02$\pm$1)$_{\rm o}$ pair come from the rim of nearby STO $\{$101$\}$ reflections. The variation in integrated intensities reveals that one dominant domain with its volume fraction more than one order of magnitude larger than the rest three domains. From the angular positions of STO (00n)$_{\rm c}$ reflections and the beam specularly reflected from sample surface, we determined that the miscut angles of the STO substrate are 0.14$^{\rm o}$ and 0.01$^{\rm o}$ along two orthogonal lateral STO $<$100$>$ directions. We defined the direction with larger miscut angle as STO [100]$_{\rm c}$, and the terrace edge is thus along STO [010]$_{\rm c}$ direction. Further analysis reveals that the dominating domain corresponds to an orientation of SRO [001]$_{\rm o}$ being aligned with STO [010]$_{\rm c}$, i.e. along surface terrace edge (Domain A in Fig. \ref{Fig5}(f)). The same dominant orientation was found in all other SRO films. The four structural domains were illustrated in Fig. \ref{Fig5}(f). Domain A and B are the 90$^{\rm o}$ domains. Domain C and D refer to the 180$^{\rm o}$ counterparts for the domain A and B, respectively. The corresponding volume fractions for the four domains are listed in Table \ref{table}. Table \ref{table} summarizes the variation of volume fraction of above four domains for three SRO films with different $t$ and ($\alpha$, $\beta$). 

\begin{table}[h]
%\begin{tabular}{h} %create 9 columns
\caption{ Thickness and ($\alpha$, $\beta$) dependents on domain volume fraction and RRR/$\rho$(5 K). $\beta$ represents the miscut direction, which is defined as the angle between the terrace edge and STO [010]$_{\rm c}$.}
\begin{tabular}{cccc} % creating 9 columns
\hline
  {\textit{t}}(nm) &($\alpha$($^{\rm o}$), $\beta${($^{\rm o}$)}) & Domain volume fraction ($\%$) & RRR/$\rho$(5 K)($\mu\Omega$cm)\\
\hline 
 26.6 & (0.14, 5) & A(92), B(6), C(1), D(1)  & 75.7/2.4\\
 4.3 & (0.08, 27) & A(37), B(21), C(18), D(24) &  11.7/24.5\\
 4.3 & (0.56, 1) & A(75), B(8), C(6), D(11) &  9.0/34.0\\
 \label{table}
\end{tabular}
\end{table}

\begin{figure}
  \includegraphics[width=\linewidth]{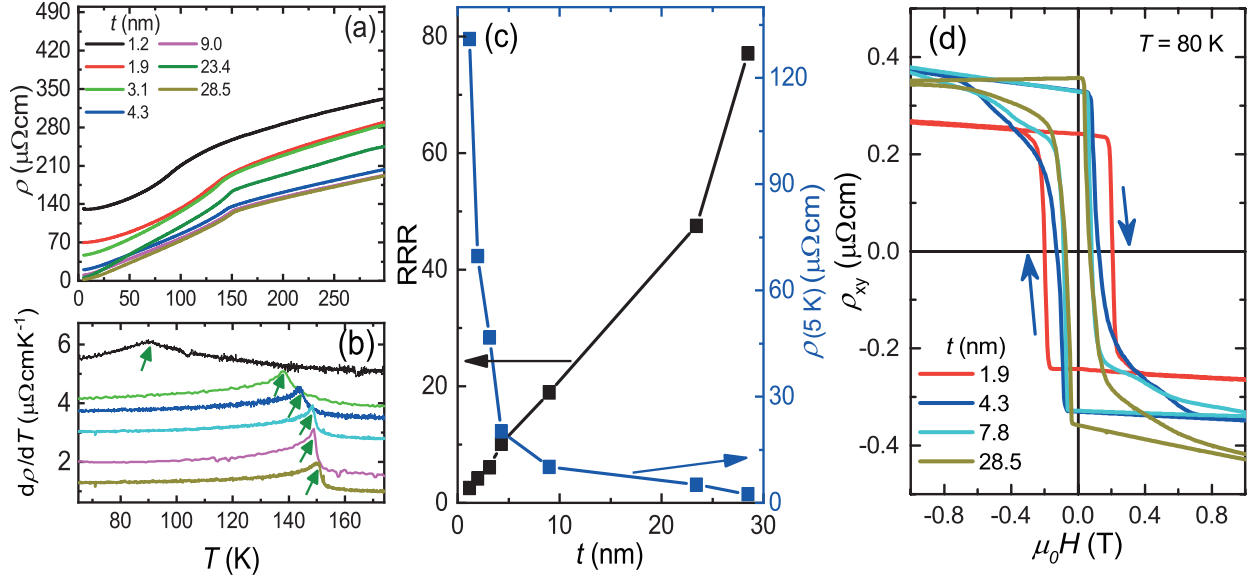}
   \caption{Transport data of SRO(\textit{t}) films on STO. (a) Temperature dependent resistivity in SRO films with different \textit{t} values. (b) Derivative of the $\rho(T)$ curves for different \textit{t} values, where the peak corresponds to the $T_{\rm c}$. (c) Variation of RRR and $\rho$(5 K), extracted from the $\rho(T)$, as a function of \textit{t}. (d) The Hall resistivity as a function of  magnetic field for different \textit{t} values at $T$ = 80 K. The presence of hysteresis loops indicates the ferromagnetic nature of the SRO films on STO.}
  \label{Fig6}
\end{figure}

We performed a thickness-dependent transport study on the films with the $\tau_{\rm IGD} = \tau_{\rm OIGD}$. Figure \ref{Fig6}(a) displays the variation of $\rho$ with temperature for different $t$. SRO films showed metallic nature down to the lowest temperature for $t>$ 2.0 nm. The $t$ $\approx$ 1.2 nm film exhibited metallic nature down to 9 K. With further reducing the temperature, a slight increase ($\approx$2$\%$) in the $\rho$ was found. A kink appeared in all the $\rho(T)$ curves due to the onset of ferromagnetism. The transition temperature ($T_{\rm c}$) was extracted from the peak location in the derivative of $\rho(T)$ curve as shown in the Fig. \ref{Fig6}(b). Figure \ref{Fig6}(c) displays the variations of the RRR and $\rho$(5 K) as a function of $t$, showing an apparent trend of increasing(decreasing) RRR($\rho$(5 K)) with growing $t$. Remarkably, the RRR rose from around 2.5 for $t$ $\approx$ 1.2 nm to around 77.1 for $t$ $\approx$ 28.5 nm. $\rho$(5 K) reduced from around 131.0 $\mu\Omega$cm for $t$ $\approx$ 1.2 nm to around 2.5 $\mu\Omega$cm for $t$ $\approx$ 28.5 nm. Figure \ref{Fig6}(d) plots the Hall resistivity $\rho_{xy}$ versus magnetic field for different $t$ values at 80 K. A typical hysteresis loop appeared in the $\rho_{xy}$-$\textit{H}$ for all SRO films with different $t$ values, providing a further evidence for the ferromagnetism in the SRO films \cite{roy2015}. 

\begin{figure}
  \includegraphics[width=14 cm]{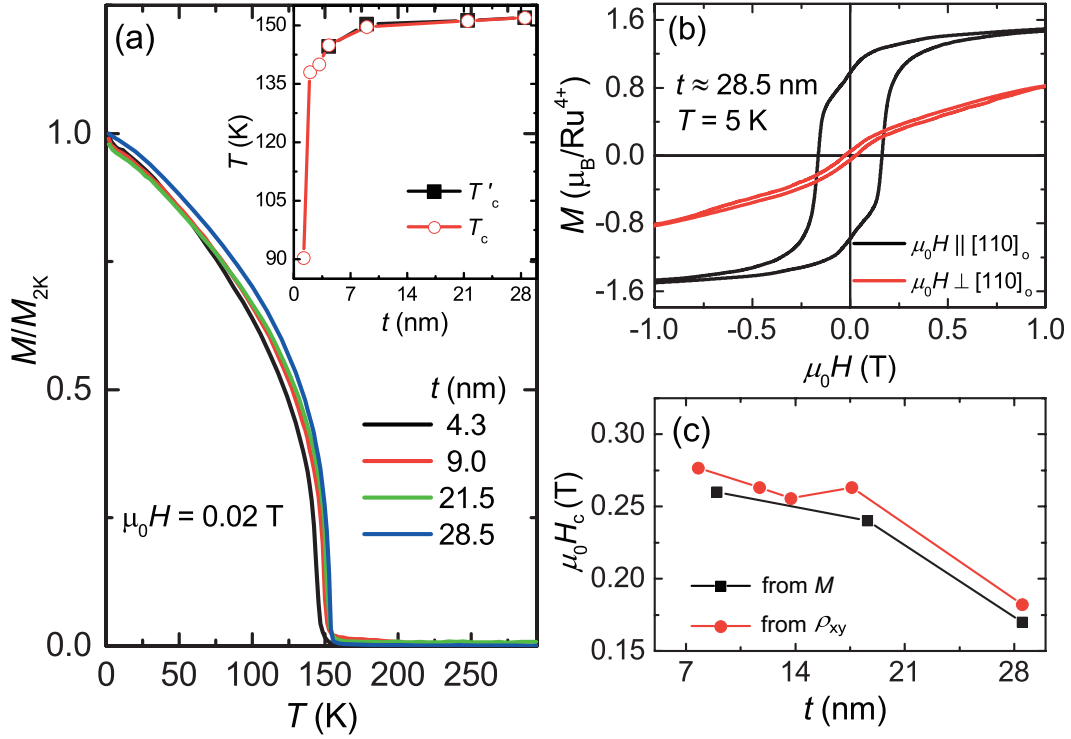}
  \caption{The magnetic properties of SRO films on STO. (a) The magnetization as a function of \textit{T} for different \textit{t }values, where the $M$ data were measured with an applied field of $\mu_0H$ = 0.02 T along SRO [110]$_{\rm o}$. Inset shows the variation of transition temperature with respect to \textit{t} extracted from the derivative of the $M(T)$ data and from the $\rho(T)$ data of Fig. \ref{Fig6}(b). (b) \textit{M-H} hysteresis loops for \textit{t} $\approx$ 28.5 nm of SRO film at $T$ = 5 K measured for $\textit{H} \parallel$ SRO [110]$_{\rm o}$ and $\textit{H} \perp$ SRO [110]$_{\rm o}$ directions, revealing a magnetic easy axis along SRO [110]$_{\rm o}$. (c) The $t$ dependent coercive field $H_{\rm c}$ extracted from $M$ and $\rho_{\rm xy}$ data with $\mu_0H \parallel$ SRO[110]$_{\rm o}$. $H_{\rm c}$ gradually decreases from about 0.28 to 0.18 T as $t$ increases from about 7.8 to 28.5 nm.}
  \label{Fig7}
\end{figure}

Figure \ref{Fig7}(a) shows the temperature dependent magnetization of the SRO films on STO, which was measured with a small magnetic field of $\mu_{0}H$ = 0.02 T along SRO [110]$_{\rm o}$. All $\textit{M}$-$\textit{T}$ curves clearly show a ferromagnetic to paramagnetic phase transition, and the $T'_{\rm c}$ was extracted from the peak location in the derivative of the $\textit{M}$-$\textit{T}$ curve. Inset of Fig. \ref{Fig7}(a) shows the variation of $T'_{\rm c}$ and $T_{\rm c}$ as a function of $t$, where the $T_{\rm c}$ values were extracted from the $\rho(T)$ curves (Fig. \ref{Fig6}(a)). The $T_{\rm c}$ of around 151 K for $t>$ 9 nm decreases to around 90 K with $t$ reducing to 1.2 nm. In the high field regime, the magnetic signal was practically linear with field strength, which was dominated by the diamagnetic background from the STO substrate. The field dependent magnetization of the SRO film can thus be obtained by subtracting a field linear component due to the diamagnetic STO, and the resulting $M$-$H$ curves for a SRO film with $t$ $\approx$ 28.5 nm at $T$ = 5 K are shown in Fig. \ref{Fig7}(b). For $H \parallel$ SRO [110]$_{\rm o}$, $M$ rapidly rises in low field regime and saturates to a value of $M \approx$ 1.5 $\mu_{\rm B}$/Ru$^{4+}$ for $\mu_{0}H \geq$ 1 T. On the contrary, for  $H \perp$ SRO [110]$_{\rm o}$, $M$ increases much slower with field strength, giving a value of $M \approx$ 0.8 $\mu_{\rm B}$/Ru$^{4+}$ at $\mu_{0}H$ = 1 T. This result reveals the presence of magnetic anisotropy with a magnetic easy axis along the SRO [110]$_{\rm o}$ \cite{kos2012,gan1998}. The coercive fields ($H_{\rm c}$) for $H \parallel$ SRO [110]$_{\rm o}$ are extracted from observed hysteresis loops in $M$ and $\rho_{\rm xy}$ at $T$ = 2.5 K. The $t$ dependent $H_{\rm c}$ is shown in Fig. \ref{Fig7}(c), exhibiting a progressive decrease from $H_{\rm c} \approx$ 0.28 T for $t \approx$ 7.8 nm to $H_{\rm c} \approx$ 0.18 T for $t \approx$ 28.5 nm.

\section{Discussion}
In earlier works, the SRO stoichiometry and oxygen vacancy driven studies have been carried out, where notable changes in the RRR, $\rho$(5 K), film's crystallinity, and $T_{\rm c}$ were reported \cite{cap2002,dab2004,lee2017,sie2007}. We grew all the films under a similar ozone environment and within the adsorption-controlled growth regime. Hence, the significant change in RRR and $\rho$(5 K) with respect to the $\tau_{\rm IGD}$ is not likely due to either the oxygen vacancy in STO \cite{sin2020} or SRO stoichiometry. To further clarify the oxygen vacancy issue, we post-annealed a sample at 400 $^{\rm o}$C for 8 hours under 1 atm O$_2$ flow, where no noticeable change in the RRR and $\rho$(5 K) was observed due to post-annealing. We also note that the rapid and monotonic increase of $\rho$(5 K) with reducing $t$ (Fig. \ref{Fig6}(d)) suggests the insignificance of the possible interface conduction channel between SRO and STO due to either atomic interdiffusion or charge transfer at the interface \cite{oht2004}.  

It was also pointed out that step-flow growth mode is more favorable for achieving atomically smooth surface in SRO thin films \cite{kos2012,hon2005,rij2004}. In a simplified model \cite{hon2005}, two relevant time scales are considered. One is the lifetime of an adatom diffusing on a terrace with width \textit{L} before being absorbed on the surface, namely $\tau_{life}$ = $L^2/2D$, and \textit{D} is the diffusion constant. The other time scale represents the time elapsed between two consecutive atoms to land on the surface, and it can be described by $\tau_{\rm land}$ = $a^2/L^2F$, where \textit{F} and $a$ are the deposition flux and the surface lattice constant, respectively. The condition for step-flow growth regime requires $\tau_{\rm life} < \tau_{\rm land}$. The additional periodicity from the initial SrO layer with a \textit{c}(2 $\times$ 2) superstructure (Fig. \ref{Fig2}(e)) not only provides ordered nucleation sites for SRO growth but also imposes a much shorter length scale as compared to the terrace width \textit{L} in the $\tau_{\rm life}$ term, which prevents the island formation and promotes the step-flow growth. As a result, the SRO film grown with the $\tau_{\rm IGD} = \tau_{\rm OIGD}$ gives rise to an excellent crystallinty and reduces(increases) the $\rho$(5 K)(RRR) by about an order of magnitude as compared to the films grown with $\tau_{\rm IGD} \neq \tau_{\rm OIGD}$.

One disadvantage for the adsorption-controlled growth technique is the random RuO$_{\rm x}$ clusters on SRO films due to the excess Ru flux during the growth process. Nevertheless, our growth condition is well within the thermodynamic growth window for an equilibrium state of SRO solid phase and RuO$_{\rm x}$ gas phase \cite{nai2018}. We thus argue that those random RuO$_{\rm x}$ clusters mostly precipitated on the film surface and did not significantly influence the crystallinity of SRO films underneath, which is supported by the observations of pronounced intensity oscillations in the CTR and off-normal L-scans on our SRO films. Moreover, RuO$_{\rm x}$ clusters are well separated by more than few microns, which is not likely to influence the transport properties of SRO. But, in the magnetization measurement, the RuO$_{\rm x}$ clusters may give an additional contribution to the background signal. 

Previous studies of SRO films on STO have revealed the presence of structural domains that is sensitive to the STO's $\alpha$ and $\beta$ parameters \cite{gan1997,wan2020}, which mostly relied on analyses of the azimuthal scans along SRO (221)$_{\rm o}$ reflections and RSMs near STO $\{$204$\}$$_{\rm c}$. Our approach by measuring the $\phi$-scan of the SRO (02$\pm$1)$_{\rm o}$ reflections allows us to probe contributions from each domain separately as demonstrated in Fig. \ref{Fig5}(h). Comparing the 4.3 nm thick SRO film grown on a nearly on-axis STO (001)$_{\rm c}$ substrate of ($\alpha$, $\beta$) $\approx$ (0.08$^{\rm o}$, 27$^{\rm o}$) with that on a largely miscut substrate of ($\alpha$, $\beta$) $\approx$ (0.56$^{\rm o}$, 1$^{\rm o}$), we noted that the volume fraction of domain A increased drastically from 37 $\%$ to 75 $\%$, as shown in table \ref{table}. The extracted domain volume fractions thus depend on ($\alpha$, $\beta$) values as expected. Moreover, for SRO films grown on nearly on-axis substrates, the dominant volume fraction increases from 75 $\%$ for $t \approx$ 4.3 nm to 92 $\%$ for $t \approx$ 26.6 nm. These results reveal a notable trend of achieving single-domain films by using STO with a large miscut angle and near axis miscut direction, i.e. small $\beta$, and increasing film thickness. 

Finally, we like to discuss the possible structural domains with SRO [00$\pm$1]$_{\rm o}$ along STO [001]$_{\rm c}$ in our SRO films. As pointed out previously \cite{kos2012,wan2020}, the magnetic easy axis for SRO films lies along SRO [110]$_{\rm o}$, and thus it gives rise to a much larger coercive field of above 1 T when the field is applied along the SRO [001]$_{\rm o}$. As demonstrated in Fig. \ref{Fig6}(d), the coercive fields in our SRO films with fields along STO [001]$_{\rm c}$ are well below 1 T, indicating a negligible population for domains with SRO [00$\pm$1]$_{\rm o}$ along STO [001]$_{\rm c}$. The conclusion is further supported by XRD results. To avoid possible overlapping with diffraction peaks from SRO (110)$_{\rm o}$ oriented SRO domains, we looked for the (221)$_{\rm o}$ reflection of (001)$_{\rm o}$ oriented SRO domains, which is located near STO (2 0 0.5)$_{\rm c}$. Only signals from STO (20n)$_{\rm c}$, where n = 0 and 1, CTRs were observed. No distinguishable peak was found within our detection limits, manifesting the negligible amount of (001)$_{\rm o}$ oriented SRO domains, if ever exist.

\section{Conclusion}
Using an oxide-MBE and adsorption-controlled growth technique, we grew SRO(\textit{t}) films on STO (001)$_{\rm c}$ and studied their thickness-dependent structural, transport, and magnetic properties. Our results revealed that within the adsorption-controlled growth regime, a control on the initial SrO growth parameters is crucial to achieve a low RR. The initial SrO layer with $\tau_{\rm IGD} = \tau_{\rm OIGD}$ results in a \textit{c}(2 $\times$ 2) superstructure, which serves as a proper template for the growth of SRO films with high-crystallinity and low RR. From thickness dependent investigations, SRO films show an orthorhombic-phase down to $t \approx$ 4.3 nm, and their metallicity and ferromagnetism were well preserved down to $t \approx$ 1.2 nm. By performing $\phi$-scan across SRO (02$\pm$1)$_{\rm o}$ reflections, four structural domains were clearly identified, comprising two 90$^{\rm o}$ domains that each has its own 180$^{\rm o}$ counterparts. The dominant domain appears to have SRO [001]$_{\rm o}$ along the terrace edge, and its volume fraction grows with increasing \textit{t}, giving a volume fraction of about 92 $\%$ for \textit{t} $\approx$ 26.6 nm. For a fixed $t$ of about 4.3 nm, the volume fraction of the dominant domain also increases from about 37 $\%$ to around 75 $\%$ when the STO miscut angle $\alpha$ increases from 0.08$^{\rm o}$ to 0.56$^{\rm o}$. Our results reveal not only the complex structural domains in SRO films but also an unexpected dictation of the film's residual resistivity and crystallinity by the initial SrO growth condition, which is crucial for the adsorption-controlled growth of SRO thin films on STO with consistent quality.

 \section{Methods}
Using an oxide-MBE technique, a series of SRO films with different $t$s were grown on the TiO$_2$-terminated STO (001)$_{\rm c}$ substrates. STO substrate was etched using aqua regia followed by annealing at 1000 $^{\rm o}$C for 12 hours in pure oxygen flow under atmospheric pressure. STO substrate was heated to a growth temperature of around 700 $^{\rm o}$C that was measured using a pyrometer. In order to avoid the oxygen loss in the STO, distilled ozone was supplied into the growth chamber whenever substrate temperature was above 150 $^{\rm o}$C. Ozone partial pressure was maintained at around 3 $\times$ 10$^{-6}$ Torr throughout the growth process. Sr and Ru was evaporated using a standard effusion-cell and e-beam, respectively. The atomic fluxes of both Sr and Ru were precalibrated using a quartz crystal microbalance. Sr flux was about 9.29 $\times$ 10$^{12}$ cm$^{-2}$s$^{-1}$ and the Ru/Sr flux ratio was kept around 2.2. The Sr-cell shutter was opened first for certain duration, and then Ru shutter was opened as illustrated schematically in Fig. \ref{Fig1}(a). The growth process was \textit{in-situ} monitored via a RHEED. A LEED was used to identify the surface atomic structure.  AFM and SEM-EDX were used to investigate the surface morphology and the film composition, respectively. High resolution X-ray scattering characterizations were carried out at beamlines TPS 09A and TLS 07A of the NSRRC, Taiwan. The transport measurements were performed using a superconducting magnet system with a variable temperature insert. The magnetic properties were studied using the commercial Quantum Design magnetic properties measurement system.

\section{Data Availability}
All the supporting data are included in the main text. The raw data and other related data for this paper can be requested from C.H.H. (chsu@nsrrc.org.tw) and W.L.L. (wlee@phys.sinica.edu.tw).
\section{Author Contributions}
A.K.S. and U.K. grew the epitaxial films and also performed magnetization measurements. B.D. and W.L.L. carried out the low temperature magneto-transport measurements and data analyses. A.K.S., U.K., S.Y., C.Y.L., and C.H.H. performed X-ray measurements and analyses. A.K.S., U.K., C.H.H., and W.L.L. designed the experiment and wrote the manuscript.  
\section{Acknowledgments}
This work is supported by Academia Sinica and the Ministry of Science and Technology of Taiwan (MOST 108-2628-M-001-007-MY3 and MOST 106-2112-M-213-006-MY3). In particular, we are grateful to H.P. Nair and D.G. Schlom for their kind sharings of valuable experience with adsorption-controlled growth technique using an oxide-MBE. 

%\section{Additional Information}
%\textbf{Supplementary Information} accompanies the paper on the physical review research website (https://XXXXX).\\ 
\textbf{Competing interests:} The authors declare no competing financial or non-financial interests.

\bibliographystyle{apsrev4-2} % for PR series journals
\bibliography{test}

\newpage 

\end{document}